\documentclass[sigconf]{acmart}
\AtBeginDocument{%
  }

\usepackage{booktabs}
\usepackage[table]{xcolor}
\usepackage{tabularx}
\usepackage{ragged2e}
\newcolumntype{Y}{>{\RaggedRight\arraybackslash}X}

\usepackage{longtable} 
\usepackage{graphicx}
\usepackage{multirow}
\usepackage{array}
\usepackage{caption}
\usepackage{enumitem}
\newlist{todolist}{itemize}{2}
\setlist[todolist]{label=$\square$}
\usepackage{arydshln}

\definecolor{rowgray}{gray}{0.97}
\usepackage{multirow}
\copyrightyear{2026}
\acmYear{2026}
\setcopyright{cc}
\setcctype{by-nc-nd}
\acmConference[CHI '26]{Proceedings of the 2026 CHI Conference on Human Factors in Computing Systems}{April 13--17, 2026}{Barcelona, Spain}
\acmBooktitle{Proceedings of the 2026 CHI Conference on Human Factors in Computing Systems (CHI '26), April 13--17, 2026, Barcelona, Spain}
\acmPrice{}
\acmDOI{10.1145/3772318.3791553}
\acmISBN{979-8-4007-2278-3/2026/04}

\begin{document}

\newcommand{\parabold}[1]{\noindent\textbf{#1.}}
\newcommand{\change}[1]{\textcolor{black}{#1}}

\title{UnWEIRDing Peer Review in Human--Computer Interaction}
\author{Hellina Hailu Nigatu}
\authornote{Both authors contributed equally to this research.}
\email{hellina_nigatu@berkeley.edu}
\orcid{0000-0001-8784-289X}
\orcid{}
\affiliation{%
  \institution{UC Berkeley}
  \department{EECS}
  \city{Berkeley}
  \country{USA}
}

\author{Farhana Shahid}
\authornotemark[1]
\email{fs468@cornell.edu}
\orcid{0000-0003-3004-7099}
\affiliation{%
  \institution{Cornell University}
  \department{Information Science}
  \city{Ithaca}
  \country{USA}
}

\author{Vishal Sharma}
\email{vishal.sharma@nd.edu}
\orcid{0000-0001-9880-5883}
\affiliation{%
  \institution{University of Notre Dame}
  \department{College of Engineering}
  \city{Notre Dame}
  \country{USA}
}

\author{Abigail Oppong}
\email{abigoppong@gmail.com}
\orcid{0000-0002-6468-1787}
\affiliation{%
  \institution{Independent Researcher}
  \city{Accra}
  \country{Ghana}
}

\author{Michaelanne Thomas}
\email{mmtd@umich.edu}
\orcid{0000-0002-2723-2157}
\affiliation{%
  \institution{University of Michigan}
  \department{School of Information}
  \city{Ann Arbor}
  \country{USA}
}

\author{Syed Ishtiaque Ahmed}
\email{ishtiaque@cs.toronto.edu}
\orcid{0000-0003-2452-0687}
\affiliation{%
  \institution{University of Toronto}
  \department{Computer Science}
  \city{Toronto}
  \country{Canada}
}

\renewcommand{\shortauthors}{Hellina Hailu Nigatu and Farhana Shahid et al.}


\begin{abstract}
Peer review determines which scholarship is legitimized; however, review biases often disadvantage scholarship that diverges from the norm. Human--Computer Interaction (HCI) lacks a systemic inquiry into how such biases affect underrepresented Global South (GS) scholarship. To address this critical gap, we conducted four focus groups with 16 HCI researchers studying the GS. Participants reported experiencing reviews that confined them to development research, dismissed their theoretical contributions, and questioned situated knowledge from GS communities. Both as authors and reviewers, participants reported experiencing the epistemic burden of over-explaining why knowledge from GS communities matters. Further, they noted being tokenized as ``cultural experts'' when assigned to review papers and pointed out that the hidden curriculum of writing HCI papers often gatekeeps GS scholarship. Using epistemic oppression as a lens, we discuss how review practices marginalize GS scholarship and outline actionable strategies for nurturing equitable epistemological evaluation of HCI scholarship.
\end{abstract}



\maketitle

\section{Introduction}
\begin{quote}
    ``\textit{We are burdened with the labor of reviewing this work. So it is not a straightforward critique of just the reviewing process, because reviewing is a labor of love and care.}'' --Study Participant (P2)
\end{quote}
Research in interdisciplinary computing venues, such as \change{the Association of Computing Machinery (ACM) conferences on Human Factors in Computing Systems (CHI), Computer-Supported Cooperative Work (CSCW), Fairness, Accountability, and Transparency (FAccT), AAAI Conference on Web and Social Media (ICWSM), and USENIX Security Symposium}, among others focuses predominantly on Western, Educated, Industrialized, Rich, and Democratic (WEIRD) contexts~\cite{Sturm-2015, Linxen-2021, Septiandri-2023, Hasegawa-2024, septiandri2024}. Authors publishing in these venues are largely based in WEIRD regions~\cite{Bartneck-2009, Sakamoto-2015, Linxen-2021, Septiandri-2023, Safir-2025, sharma2025sustainability}. Even research in non-WEIRD contexts is frequently produced by researchers affiliated with WEIRD institutions~\cite{giwa2015insider, sharma2025sustainability}.

\change{Originating in psychology~\cite{henrich2010weirdest} and later adopted in Human-Computer Interaction (HCI)~\cite{Linxen-2021}, the WEIRD framing reveals that most human studies in these disciplines rely on WEIRD samples, who represent only 12\% of the global population. Although related frameworks such as Anglocentrism and Coloniality capture Western hegemony in knowledge production, in this paper, we adopt the WEIRD framing because it foregrounds how epistemic norms codify knowledge based on Western populations as universal, shaping whose perspectives are legitimized in HCI research. This makes it particularly suitable for interrogating systemic inequities that arise not just from geopolitical dominance but from disciplinary practices that normalize WEIRD contexts.} 



\change{We examine how current epistemic norms impact scholars from the Global South (GS).\footnote{Following~\citet{Schneider-2017}, we recognize that while the term Global South has been mobilized to expose injustices and dependencies, without careful conceptualization it risks reinforcing North–South dichotomies, reproducing inequality, and reifying global hierarchies similar to the earlier notion of the “Third World”~\cite{harding2016latin, pinheiro2024third}. Therefore, we use these terms with care, directing our attention to historical and socio-political power dynamics rather than normative geographic labels. We distinguish between settings historically advantaged by systems of colonialism, imperialism, and capitalism, and those that have been systematically marginalized or exploited by such structures~\cite{harding2016latin,remy2018evaluation,mohanty2003under}. In doing so, we also acknowledge the coexistence of pluralist conceptualizations, such as Global Majority and Majority World~\cite{alam2008majority} that our participants also used.} Although traditionally the term GS refers to the geographies of economically developing nations, we use it to center communities with ``shared experiences of colonialism, extraction, and resistance across the globe~\cite{radiya2025same}.'' Among diverse non-WEIRD contexts, we focus on the GS because it exemplifies how epistemic inequities manifest across geopolitical, economic, and institutional dimensions. Structural barriers, such as limited research infrastructures in the GS~\cite{collyer2018, dougan2023}, citational injustice~\cite{Kumar-2021}, the pressure to publish in English which is not the first language in most GS countries~\cite{smith2014,bar2012editors}, and the undervaluing of interdisciplinary research in the GS~\cite{sharma2025sustainability}, constrain epistemological contributions from these communities. Thus, the GS offers a salient counterpoint to WEIRD framing, illustrating how historical power relations shape whose knowledge is ratified in computing research.}

\change{We specifically examine} the peer review process, which controls what scholarship is legitimized, published, and circulated~\cite{mondal2023peer,mercer2018value,morley2021now,lee2013bias}. The review process in most HCI venues is \change{double-anonymized}, masking the identities of authors and reviewers to promote impartiality. Yet, biases often persist through paternalistic comments towards non-Western populations~\cite{Kumar-2021}, assumptions that grammatical errors by authors \change{whose first language is not English reflect a lack of scientific rigor}~\cite{Lepp-2025}, and the imposition of Western-centric ethical standards on GS research studies~\cite{De-2025}--all of which reinforce a form of WEIRD gaze. \change{However, there is little systemic inquiry into why these WEIRD reviewing dynamics prevail, how they impact authors studying GS contexts, and the challenges that reviewers themselves face when reviewing GS scholarship.} To address this critical gap, in this study, we ask:


\textbf{RQ1:} How do WEIRD reviews affect GS scholarship in HCI?

\textbf{RQ2:} What challenges do HCI researchers face when reviewing GS scholarship, and how do they navigate them? 

\textbf{RQ3:} How might we nurture a more equitable peer review process that is specifically inclusive of the GS scholarship in HCI?

To answer these questions, we conducted four focus group discussions with 16 HCI researchers, who \change{engaged in GS scholarship either as authors or reviewers. Throughout the paper, we use the term GS scholarship to refer to (a) research produced by scholars from the GS, regardless of topic, and (b) research focused on GS populations, regardless of the researcher’s geographic location.}
Our findings reveal that WEIRD reviewing dynamics are pervasive in HCI research, often going beyond paternalistic or dismissive comments. WEIRD dynamics frequently limit GS scholarship to development research, disregard theoretical contributions from GS communities, and question authors' expertise based on positionality statements and monolithic assumptions about GS contexts.
Participants from the GS highlighted that when called upon to review GS scholarship, they are often reduced to tokenized \textit{``cultural experts.''} \change{They reported experiencing \textit{``epistemic burden''} when defending 
cultural contexts to fellow reviewers}.
We also noted that scholars from the GS often reproduce WEIRD reviewing practices, 
underscoring the systemic nature of the issue.
Based on their lived experiences, participants outlined several strategies for authors, reviewers, and conferences to systematically address the WEIRD gaze when evaluating research from the GS. These include allyship from scholars with relative privileges, deliberate effort from reviewers and paper chairs in countering biased reviews, and systemic reforms within HCI to prevent discriminatory practices.

As embodied in P2's quote at the beginning of this section, \change{our aim is not to criticize individual reviewers but to examine how structural and epistemic conditions shape the evaluation of GS scholarship}. We take a reflective, action-oriented approach and leverage Dotson's \textit{epistemic oppression}~\cite{dotson2012cautionary, dotson2014conceptualizing} as a theoretical lens \change{because it provides a granular, multi-level analysis of knowledge exclusionary practices and outlines pathways for systemic change. By analyzing epistemic exclusion across three critical dimensions: first-order (compromising scholars' credibility), second-order (limiting access to shared epistemic resources), and third-order (structurally preventing certain knowledge perspectives from gaining legitimacy)--this lens assists us in examining how power dynamics and Western-centric epistemic traditions embedded in the HCI peer review process devalue GS scholarship. Crucially, it allows us to move beyond critiques to propose concrete strategies for transforming WEIRD review practices, thereby offering a constructive pathway toward more equitable knowledge production.}


Our paper makes two key contributions. First, we present a qualitative study that centers the peer review experiences of both authors and reviewers of GS scholarship and uses epistemic oppression as a lens to examine how Western epistemic hegemony shapes current review practices in HCI. 
Second, moving beyond diagnosis, we offer concrete recommendations grounded in participants’ insights and Dotson’s \cite{dotson2014conceptualizing} framework of first-, second-, and third-order changes, involving diverse actions and strategies at individual, community, and systemic levels. 
This paper extends the concept of epistemic oppression to the peer review process, which currently functions as a gatekeeping mechanism, and suggests steps toward \textit{unWEIRDing peer review in HCI}. We conclude this paper with an invitation to the broader Special Interest Group on Computer-Human Interaction (SIGCHI) community to join us in making the peer review process more equitable and welcoming to GS scholarship.




\section{Background \& Related Work}
We first situate our work by providing an overview of the existing peer-review process, both within and outside HCI. Then, we describe the lens of epistemic oppression, which we use to contextualize our findings and suggest ways to make the peer review process more inclusive and equitable.

\subsection{Peer Review in HCI \& Beyond}

Peer review is widely regarded as a cornerstone of academic knowledge production. It functions as a quality assurance mechanism~\cite{lee2013bias} or a ``system of institutionalized vigilance''~\cite{merton1973sociology}, determining what research enters the scholarly record. In the field of HCI and the broader computing discipline, peer review evaluates the perceived validity, novelty, and rigor of research while shaping the intellectual trajectory by privileging certain topics, methods, and epistemologies~\cite{lee2013bias,rogers2020can,sun2022does,crane2018questionable}. This evaluative process carries high stakes: reviews directly influence researchers’ career trajectories, access to funding, and professional recognition, with particularly acute consequences for those from historically marginalized regions or underrepresented institutions~\cite{snodgrass2006single,rogers2020can} (see also~\cite{Kumar-2021, Lepp-2025, De-2025}). 

Although peer review aspires to fairness and objectivity, prior research within HCI~\cite{reynolds2022reviews,Kumar-2021, De-2025, Lepp-2025, soden2024evaluating} and beyond~\cite{rogers2020can,smith2010classical,crane2018questionable} (see also~\cite{haffar2019peer,stelmakh2019testing,lee2013bias}) has identified persistent biases that undermine these ideals. These include prestige bias (favoring work from well-known institutions), nationality bias (favoring authors from the same country where the journal or conference is based), and language bias (favoring manuscripts written by native English speakers)~\cite{lee2013bias,bender2019benderrule,rogers2020can,Lepp-2025}. Even with the \change{double-anonymized} review process
---a practice designed to mitigate the biases---reviewers may still infer an author's identity or institutional affiliation from writing style, research context, or citation practices, allowing these biases to creep back into the review process~\cite{snodgrass2006single,rogers2020can,sharma2025sustainability}. 

Particularly in HCI, where submissions range from ethnographic studies to technical system designs, review biases often intersect with methodological preferences, leading to uneven treatment of research that diverges from dominant paradigms~\cite{soden2024evaluating}. 
Several studies have noted manifestations of biased reviews in HCI: through paternalistic or dismissive comments that question the relevance, rigor, or novelty of research in GS contexts~\cite{Kumar-2021}; assumptions that grammatical errors or non-standard English reflect poor scientific quality rather than diverse linguistic background~\cite{Lepp-2025}; pressure to include GS locations in paper titles~\cite{Kou2018}; and the imposition of Western-centric ethical standards that may be inappropriate or infeasible in non-WEIRD settings~\cite{De-2025}. 

However, there is little systematic understanding of \change{why these dynamics exist and how they impact} GS scholarship. To address this critical gap, we examine the dual experiences of both the authors and reviewers of GS scholarship, exposing how they navigate power-laden reviewing dynamics and articulating pathways for more equitable practices in HCI and beyond. 

\subsection{The Lens of Epistemic Oppression}

Epistemic injustice describes the ethical harms that occur when individuals or communities are silenced, dismissed, or misrepresented in their capacity as knowers and contributors to knowledge~\cite{fricker2007epistemic}. It draws attention to identity-based prejudices that limit marginalized groups' ability to participate in and shape information ecosystems~\cite{kay2024epistemic}. Researchers have traced its roots to structural forces such as colonialism, imperialism, whiteness, and patriarchy, which systematically shape whose knowledge is recognized and ignored~\cite{spivak2023can,fricker2007epistemic,langton2010epistemic}. For example, Spivak~\cite{spivak2023can} notes how colonial systems marginalized and erased local knowledge traditions in the GS; Collins~\cite{collins2022black} highlights how Black women in the United States have historically been denied credibility through controlling stereotypes and cultural devaluation. Building on such work, theorists have distinguished between different forms of epistemic injustice~\cite{langton2010epistemic,fricker2007epistemic,nikolaidis2021third}: \textit{testimonial injustice}, when bias or prejudice undermines the credibility of a speaker; \textit{hermeneutical injustice}, when dominant interpretive frameworks exclude or obscure certain experiences; and \textit{contributory injustice}, when the insights and interpretive resources of marginalized groups are systematically excluded from shaping collective understanding. In HCI, researchers have employed epistemic injustice to surface how marginalized researchers are systematically devalued or excluded. This includes work highlighting the challenges faced by scholars---who are either Black women~\cite{erete2021can}, have disabilities~\cite{ymous2020just}, based in the GS~\cite{Kumar-2021}, or work with marginalized groups~\cite{ajmani2023epistemic,ajmani2025moving}.

Dotson~\cite{dotson2012cautionary,dotson2014conceptualizing} extends the epistemic injustice framework through the concept of epistemic oppression. She refers to epistemic oppression as ``persistent epistemic exclusion that hinders one's contribution to knowledge production''~\cite[p.~115]{dotson2014conceptualizing}. Instead of the isolated acts of testimonial or hermeneutical injustice, epistemic oppression points to systematic patterns that silence marginalized knowers and reinforce their exclusion over time. By situating epistemic injustice within broader histories of oppression, Dotson urges us to see how knowledge production itself can be complicit in sustaining inequity, while also pointing toward the need for transformative practices that actively resist such exclusion~\cite{dotson2014conceptualizing}. She distinguishes between \textit{first-order, second-order}, and \textit{third-order} epistemic exclusion~\cite{dotson2014conceptualizing}: first-order exclusion occurs when individuals are prevented from contributing knowledge due to credibility deficits; second-order exclusion arises when marginalized groups lack access to shared epistemic resources or are denied participation, and; third-order exclusion operates when dominant epistemic systems themselves are structured to perpetuate ignorance, ensuring that epistemic tools from certain groups remain excluded across time and institutions. 

To counter epistemic oppression, Dotson~\cite{dotson2014conceptualizing} outlines three degrees of change: first-, second-, and third-order. First-order change focuses on addressing problems as they emerge while leaving the underlying structures and norms intact. Second-order change involves shifting the values and practices of individuals or groups, fostering new strategies, perspectives, and ways of acting that improve the system’s functioning. Third-order change goes further by interrogating and potentially transforming the deep social imaginaries that sustain and legitimize the prejudiced status quo. Together, these layered interventions remind us that justice requires fixing immediate harms while reimagining the cultural and structural conditions that enable oppression in the first place.

We leveraged the epistemic oppression framework to draw attention to how structurally biased systems of academic publishing and peer review undermine the epistemic agency of underrepresented GS scholars in HCI. We critically examine how reviewers currently engage with GS scholarship and how bias shapes the credibility afforded to such work. Further, we leveraged the framework to suggest first-, second-, and third-order changes that can foster fairer, more inclusive review practices in HCI, rippling outwards towards the broader computing discipline and beyond. 

\section{Methods} \label{sec:methods}

We conducted four focus groups with 16 participants. Our study was approved by the Human Subject Review Board at two institutions where the co-first authors are affiliated. The co-first authors collected and analyzed the data. 

\begin{table*}[!ht]
\caption{Distribution of HCI peer reviewing experiences among initial respondents and focus group participants.}
\label{tab:review_exp}
\centering
\begin{tabular}{@{}lcc@{}}
\toprule
\textbf{Reported Experiences} & \textbf{Initial Respondents} & \textbf{Focus Group Participants} \\
 & N=30 (\%) & N=16 (\%) \\
\midrule
Has published HCI papers on GS contexts & 25 (83\%) & 15 (94\%) \\
Has reviewed HCI papers on GS contexts & 28 (93\%) & 16 (100\%) \\
Has received biased reviews related to GS contexts & 13 (43\%) & 10 (63\%) \\
Has received biased reviews targeting author's identity in relation to GS & 9 (30\%) & 7 (44\%) \\
Is from the GS but does not study or review research on GS contexts & 4 (13\%) & -- \\
\bottomrule
\end{tabular}
\end{table*}

\subsection{Participant Recruitment} To recruit participants, we distributed a screening survey via social media, such as LinkedIn, Twitter/X, and BlueSky, and used our professional networks. 
\change{A total of 30 researchers responded to our survey. Since our research questions center on how WEIRD reviewing dynamics impact GS scholarship, we invited those who met at least one of the following criteria: (1) had published HCI papers on GS contexts, (2) had reviewed HCI research related to GS contexts, or (3) as authors, had encountered biased reviews from HCI venues targeted toward their identity or research focus on the GS. This recruitment strategy enabled us to capture perspectives from both authors and reviewers, since authors often lack full visibility into how reviewers and associate chairs (ACs) make decisions. Further, by directly involving HCI researchers who were either from the GS or had published or reviewed HCI research about the GS, we followed the best practices of engaging the affected communities when conducting research about them~\cite{Yiwei2019Nov, eslami2015Apr}.} 

In total, 16 HCI researchers participated in our study. \change{Table~\ref{tab:review_exp} and \ref{tab:focus_grp} summarize their review experiences and demographic details. Participants represented different career stages, including doctoral researchers (n=8), postdoctoral researchers (n=3), professors (n=2), and industry practitioners (n=3). Their research spanned diverse GS contexts, such as Central and South Asia, Africa, and Latin America. However, the majority of them were affiliated with Global North institutions. 
Two participants were not from the GS but had extensive experience of conducting research in GS contexts.} 

\change{On average, participants had 8 years of HCI research experience (SD = 5.6). All had served as peer reviewers for major HCI venues (e.g., CHI, CSCW), and six had additionally served as ACs. Around two-thirds of the participants who had published GS-related work in these venues reported receiving biased reviews, either directed toward the populations they studied or toward their own identities. By including both authors who had encountered such biases and those who had not, our sample allowed us to examine perceived inequities in HCI peer reviews from different angles.}

\begin{table*}[!ht]
\caption{Details of focus group participants. Specific country names have been omitted from institutional affiliations to protect participant anonymity.}
\resizebox{.8\textwidth}{!}{
\centering
\begin{tabular}{lllll}
\hline
\textbf{PID} & \textbf{Career Stage} & \textbf{Years in HCI} & \textbf{Research Context} & \textbf{Institutional Affiliation} \\ \hline
P1  & Industry practitioner    & 11+ & India            & North America \\
P2  & Professor                & 8+  & Nepal            & North America \\
P3  & Doctoral researcher      & 10+ & Pakistan         & North America \\
P4  & Post-doctoral researcher & 8+  & Bangladesh       & North America \\
P5  & Post-doctoral researcher & 6+  & India            & North America \\
P6  & Doctoral researcher      & 1+  & Kazakhstan       & North America \\
P7  & Doctoral researcher      & 7+  & Kenya, Uganda    & North America \\
P8  & Post-doctoral researcher & 9+  & Bangladesh       & North America \\
P9  & Industry practitioner    & 26+ & Africa, India    & Africa \\
P10 & Professor                & 6+  & Uganda           & North America \\
P11 & Doctoral researcher      & 4+  & India            & North America \\
P12 & Industry practitioner    & 8+  & Colombia         & North America \\
P13 & Doctoral researcher      & 3+  & Nigeria          & North America \\
P14 & Doctoral researcher      & 5+  & Bangladesh       & Europe \\
P15 & Doctoral researcher      & 7+  & Nepal            & North America \\
P16 & Doctoral researcher      & 9+  & India            & South Asia \\
\hline
\end{tabular}
}
\label{tab:focus_grp}
\end{table*}

\subsection{Data Collection and Analysis} We conducted focus groups to provide a shared space for GS researchers, who often lack formal avenues to address concerns about biased reviews and typically navigate these challenges in isolation. This format enabled authors--whether or not they had biased review experiences--and reviewers to listen to, build on, and respectfully contest each other’s perspectives~\cite{acocella2012focus}. This dynamic helped participants recognize their own experiences in others' accounts and collectively articulate phenomena they might not have had the language to describe on their own. Moreover, focus groups are well-suited for research examining power dynamics and marginalization, as the group setting can reduce the power differential between researchers and participants while fostering solidarity among those with shared experiences~\cite{bosco2010focus,freeman2006best}.

All focus groups were in English \change{because it was the only shared language across our multilingual research team and participants, most of whom were affiliated with Global North institutions. We acknowledge that this choice sits in tension with our critique of the hegemony of English in academic knowledge production~\cite{bar2012editors}. Our research process is thus bounded by the same structural conditions that we interrogate in this paper. Following prior work~\cite{tietze2013victorious}, we therefore treat this tension as a site of hegemonic struggle by drawing attention to and reflecting on how English shaped both participants’ and our own contributions.} 

We conducted the focus groups online via Zoom. Each focus group lasted for around 60 minutes and included 3--5 participants. Participation in the focus group was voluntary. 
Before joining the focus groups, participants signed an informed consent form. We also collected verbal consent from each participant before recording the session. Participants were given the choice to turn off their video or mute their microphones when not speaking if they preferred not to have their video or background audio included in the recording. 

The co-first authors facilitated the focus groups. We 
debriefed the participants about the goal of the focus group. 
We structured our focus group into three sub-sections. First, we asked participants about their experiences receiving reviews \change{from HCI venues}, and many reflected on biased reviews that targeted their identity as a GS researcher or the communities they worked with in the GS. Then, we asked participants to share their experiences of reviewing \change{HCI} research, either from or about the GS, and to reflect on the tensions and challenges they encountered as reviewers. Finally, we asked participants to reflect on the way forward, during which they provided practical strategies for handling \change{systemic issues in the peer review process} and highlighted best practices in peer review from other disciplines. 

We recognize that inviting GS scholars to reflect on biased reviews may inadvertently place additional burden on those navigating systemic inequities within academia. Our intention is not to ask participants to resolve their own oppression, but rather to document and foreground their experiences, which are often overlooked within broader critiques of the peer review process. 

We continued inviting participants for focus groups until their responses reached theoretical saturation~\cite{pandit-1996}. After each focus group, the co-first authors independently conducted open coding on focus group transcripts. They met regularly to synthesize initial themes following reflexive thematic analysis~\cite{braun-2006} and compared the emerging themes against those identified in prior sessions. By the fourth focus group, while participants described context-specific manifestations of WEIRD reviews, their accounts consistently mapped onto the same overarching pattern. Saturation was determined when subsequent focus groups yielded experiences that fit within existing thematic categories. In the end, our sample comprised 16 participants across four focus groups, consistent with prior evidence that three to six focus groups are typically sufficient to achieve theoretical saturation for focused research questions~\cite{guest2017many}. 
In total, we collected nearly four hours of audio recordings, 17 pages of detailed notes, and 103 pages of transcripts. Our final analysis produced 438 unique open codes, grouped into 3 first-level themes and 13 second-level themes.

\subsection{Positionality} All authors of this work come from diverse interdisciplinary computing backgrounds and have roots in the GS. Our research is situated in and informed by diverse GS contexts, including Africa, Latin America, and South Asia. This collaboration emerged from our shared experiences of grappling with WEIRD reviews in HCI venues and recognizing the lack of formal spaces to systematically address such concerns beyond informal conversations and social media exchanges. Since most of the authors are affiliated with GN institutions, we acknowledge our privilege of having access to greater epistemic resources when writing for HCI venues compared to our peers in the GS. Additionally, as reviewers and associate chairs (ACs) of HCI venues ourselves, we recognize the power we hold to normalize and perpetuate review practices that often disadvantage GS scholarship. In undertaking this work, we adopt a reflexive ``nervous'' stance that resists the position of the distanced, powerful researcher~\cite{shankar2022brown}. We try to attune to the embodied, affective, and structural ways in which we, as reviewers, are implicated in the system--ingenuously reinforcing dominant assumptions about what counts as credible knowledge. In doing so, we aim to go beyond simplistic South/ North binaries in knowledge production to examine the systemic factors that contribute to biases in peer review, while acknowledging the voluntary labor and community service provided by reviewers. We hope to contribute to the emerging conversations in HCI on changing the status quo (e.g.,~\cite{sharma2025sustainability,Kumar-2021,sharma2023post}) toward building a truly inclusive community that nurtures knowledge and diverse perspectives from the GS. 

\section{Findings} \label{sec:findings}
We examine the experiences of HCI researchers as they navigate the dual role of authors and reviewers in the current peer review system. First, we present participants’ experiences of receiving reviews on work involving GS communities (\ref{sec:receiveing}). 
Then, we discuss the challenges and tensions participants faced when reviewing research from or about the GS (\ref{sec:giving}).
Finally, we draw out the strategies that participants reported to improve the peer review process to support GS scholarship (\ref{sec:forward}).

\subsection{Perceived Bias in Reviews}\label{sec:receiveing}
\change{Our participants reported receiving mixed reviews depending on the venue. Few participants reflected that smaller venues such as the Participatory Design Conference and Data \& Society workshops were more receptive to diverse scholarship from the GS compared to larger venues, such as CHI and CSCW. While some shared that they often received positive reviews from CHI and CSCW, nearly two-thirds of the participants reported experiencing negative reviews either targeted towards their identity or GS research context (refer to Table~\ref{tab:review_exp}). Below, we present how different participants perceived and experienced biased reviews in relation to \textbf{RQ1}.}

\subsubsection{What Do WEIRD Reviews Look Like?} Participants felt the \change{negative} reviews they \change{often} received were written from a WEIRD lens, offering little constructive feedback. They expressed that the reviews frequently criticized the novelty of their work, made problematic remarks, and imposed a double standard on GS scholarship. 

\paragraph{Novel Contribution} Participants studying socio-technical issues in the GS were repeatedly asked to justify why their research sites were novel or interesting. P9 commented that GS researchers always have to answer \textit{``why should one care about technology usage in Nigeria or Kenya but never for the US and Europe.''} Participants pointed out that while problems such as climate change, machine translation, and gender-based violence were well-studied in Western contexts, the lack of pluralistic perspectives from the GS was often overlooked in reviews, dismissing their research in these contexts as less novel. P4 commented:
\change{
\begin{quote}
\textit{``I think as Global South scholars we have this extensive pressure that we always have to bring something new `exotic' for the Global North audience to recognize the knowledge we are producing.''}
\end{quote}}

Participants described feeling pressured to highlight novelty, which created tension about how to foreground community needs, especially when those needs were also observed in Western contexts. They felt compelled to justify why knowledge from these communities mattered. P2 drew on ~\citet{Pierre-2021}'s notion of \textit{``epistemic burden''} and emphasized that the pressure to educate others by portraying the GS as exotic placed a similar burden on them.
In contrast, they felt similar justifications were rarely expected of research in Western contexts. P5 shared:
\begin{quote}
    \textit{``I study socio-politically disturbed context in India, where mainstream assumptions and technological solutions do not hold. When I propose designs that are low-tech and not fancy but suitable to the community, the reviewers ask how it is technologically novel, which I find weird because we also need to account for such social contexts.''}
\end{quote}

P1 and P3 explained that to satisfy such reviews, they have to \textit{``over-explain''} the socio-cultural contexts of the GS, which particularly burdened them when writing for ACM venues like FAccT and AIES that imposed strict page limits. Additionally, P6 pointed out that the lack of prior research in largely understudied GS contexts, \change{such as Kazakhstan} created \textit{``a vicious cycle''}, as reviewers often demanded citations to validate situated knowledge that was yet to be documented. P1 narrated:
\begin{quote}
    \textit{``My reviewers asked why I cited news articles to justify studying caste among Muslims in India. They think caste is only a Hindu concept, but that's not how it works in India$\dots$and there's not much academic work on this topic. Eventually, I had to find a reference to a government website to prove my point.''}
\end{quote}

These experiences illustrate that Western-centrism in knowledge production often invalidates situated knowledge from the GS and undermines research contributions focusing on these communities.

\paragraph{Problematic Remarks} Participants reported receiving reviews that were biased against their scholarly identity in relation to the GS. 
Participants who were from the GS expressed frustration at how often reviews asked them to have their work checked by a native English speaker, undermining their contributions based on grammar or writing style rather than research rigor. Others recounted receiving paternalistic comments toward the GS contexts they studied. For example, P14 shared that one of the reviews dismissed their critical AI scholarship in the GS by insisting that \textit{``the AI revolution can uplift the marginalized Global South community.''} \change{They pointed out that such comments overlook the fact that GS communities do have agency and may not consider themselves as vulnerable. P9 further pointed to the differential treatment of HCI research from different GS contexts:
\begin{quote}
    \textit{``The reviewer asked why would small businesses in Africa want to use generative AI? But they're} [Africans] \textit{using generative AI. There's these implicit biases in how people think about business and work in African context. I think it's even worse than when I used to work in India. Since there's a bigger HCI community from India, there is more acceptance for research based in India.''}
\end{quote}
}
Participants often received reviews that reflected uninformed, monolithic assumptions about the GS. 
P3 stated that they were asked to justify why their study in Pakistan did not examine caste, as the review insisted that any study focusing on South Asia must address caste. Similarly, P8 expressed that their research on the Hindu religious minority in Bangladesh was challenged on the grounds that Hinduism is the dominant religion in South Asia. P10 reported receiving reviews questioning why their work did not examine a particular African language, even though the community they studied in Uganda did not speak it. Participants felt such reviews disregarded their agency as knowledge producers about the community to which they belonged. P1 said that responding to such comments created undue labor for GS researchers, since they could be easily avoided if reviewers considered doing \textit{``a simple Google search''}. Thus, monolithic assumptions inflicted additional burden and undermined the scholarly agency of GS researchers due to their identity and engagement with GS communities. 

\paragraph{Bucketing GS Scholarship} Our participants reflected on review experiences, where they felt their work was not treated as legit enough until they studied specific topics. They stressed that reviews that judged a researcher's expertise solely based on the geography could be harmful. P9 shared facing pushback in reviews when doing non-developmental research, \change{such as workplace or human-AI interaction studies in Africa}. P7 echoed a similar concern, saying:
\begin{quote}
    \textit{``I do lots of theory research. In my recent experience, it seems like we've been provincialized. When reviewers figure out you come from these parts of the world} [GS],  \textit{you are expected to write about health, education...but not the intellectual theory building work.''}
\end{quote}

Participants who drew on critical scholarship from the GS were often asked to justify why they did not apply theories developed in Western contexts. They felt forced to cite Western frameworks first, before engaging with GS scholarship, even when the lenses from the GS were more relevant to their research. P1 expressed their frustration, narrating:
\begin{quote}
    \textit{``The history didn't begin with the ship landing in Boston. When I'm adopting a lens based on the Indian context that I'm studying, it doesn't invalidate existing Western frameworks. Two things can co-exist, but the reviews barely acknowledge that.''}
\end{quote}

These instances reveal how current review practices reinforce Western epistemic hegemony, marginalizing theoretical frameworks and lenses from the GS.

\paragraph{Differential Treatment of GS Research}
Our participants felt burdened by the expectation to include the name of the studied country in their paper titles. They pointed out that studies in Western contexts were rarely expected to name the geography, while knowledge from GS communities was often treated as parochial. 
Participants also reported mixed experiences with including positionality statements in their papers. P13 felt that without such a statement, reviews tended to question their expertise in conducting research in the GS, \change{forcing them to disclose personal details as a way to pre-empt skepticism. P9 further highlighted the disproportionate burden of including a positionality statement:
\begin{quote}
    \textit{``We always get asked what's the positionality of the authors, and that would be fine if they} [reviewers] \textit{asked that of people doing quantitative studies in the US. But they don't. Even if you're doing a qualitative study in the UK, nobody asks you about your positionality. It's somehow your positionality only matters when you're doing work for the Global Majority.''}
\end{quote}}

Participants using qualitative and ethnographic methods to study GS contexts faced disproportionate scrutiny during peer review and were often asked to justify their methodology. P14 shared:
\begin{quote}
    \textit{``I think what kind of method we will use in our research is a very ontological and epistemological question. For the types of problems we study in Global South contexts, it is often more suitable to undertake ethnographic approach. But while publishing in elite HCI venues, I have seen they are more convinced when we come up with a ``data-driven'' approach and have lots of data to prove our point.''}
\end{quote}

These examples demonstrate how differential treatment of GS research places an additional burden on HCI researchers doing community-centered work in the GS. 



\paragraph{Generalizability} Participants highlighted a lack of recognition for contextually grounded work in the current review process, requiring them to constantly justify the generalizability of their findings. Even when they explicitly presented their work in the GS as a case study, reviewers still demanded generalizing their findings. P9 felt such reviews imposed \textit{``invisible labor''} on GS scholars if they decided to argue why generalizability was not always needed. P1 and P3 stated that, previously, reviews required papers from the GS to demonstrate universal applicability. However, they observed that the expectations were now extended to all low-resourced GS contexts, as P1 posited:
\begin{quote}
    \textit{``They} [reviewers] \textit{have stopped asking what about the US or UK. Now the reviews are like how do your findings generalize to other regional languages from India. For long we have argued that we need to study diverse Global South contexts. These reviews feel like a passive weaponization of the same argument. India alone has hundreds of languages. It's not feasible to conduct experiment in many of them or generalize findings due to lack of resources and technologies.''}
\end{quote}

Reviews often expect researchers to conduct large-scale studies and collect lots of data to prove the generalizability of their findings. For example, P16 observed that reviews in major ACM conferences often expressed skepticism about research on algorithmic bias in low-resource contexts due to limited availability of data from GS regions. P4 elaborated further:
\begin{quote}
    \textit{``I usually get this review: `you haven't tested your work with hundreds of participants. how do you know it works?' People have this universal assumption that data-intensive approach that works in the Global North will work in the Global South as well. They don't realize that first, we need to figure out what works for these communities, what will be the proper course of action in low-resource contexts, what are the fundamental differences before running 1000--10,000 participants study.''}
\end{quote}

\change{On the contrary, P3 and P8 shared that when they provided design recommendations based on their studies in Pakistan or Bangladesh, reviews often dismissed those recommendations, citing they would not be generalizable because they were based on GS communities .} These experiences reveal that GS researchers face both unjust pressure and disparate reviewing criteria to establish the generalizability of their findings. Whereas, results from Western contexts are often assumed to be universal~\cite{baber2003provincial, cheon2020usa}. These instances exemplify the multifaceted and deeply entrenched nature of Western-centrism in the peer review process, undermining HCI scholarship from the GS.

\subsubsection{\change{How and Why WEIRD Reviewing Dynamics Manifest?}}
Early-career researchers admitted that despite receiving WEIRD reviews, the reputational capital of big venues like CHI and CSCW compelled them to submit their work there. P7 expressed concern that if they published their work in more receptive but localized venues like AfriCHI, their HCI scholarship might not be treated seriously. P5 and P6 stressed that even when presented in politically correct language, such reviews often felt arbitrary and demoralizing, leaving them uncertain about the value of their work and place in the HCI community. Participants stressed that even if such reviews were unintentional, they were harmful. They highlighted that inequitable incentives and power dynamics made it difficult for them to resist WEIRD reviews, since paper acceptance was crucial for launching their careers. As a result, junior researchers felt powerless and pressured to comply with reviewers’ requests, even when those requests seemed unjust or biased. P3 commented:
\begin{quote}
    \textit{``Although it's called peer review, there's this hierarchy and power differential between reviewers and authors. They} [reviewers] \textit{have the power to make or break your work. It's often tricky how to disagree politely if you want to retain your intellectual argument. In my experience, when we reached out to the reviewers and 1AC about their comments, they took it negatively.''}
\end{quote}


When asked, participants explained why they think WEIRD assumptions persist in the review process; many acknowledged the randomness in reviewer assignments. They mentioned that WEIRD reviews were symptomatic of the broader decline in review quality as HCI conferences grew larger. As an associate chair (AC) of HCI venues, P2 noted that reviewers were often overburdened, leading them to be unable to adequately engage with the paper. Participants believed that if the ACs did not understand the context, the work would suffer in the review process. In their opinion, the shortage of expert reviewers familiar with GS contexts primarily led to uninformed reviews. 

\change{Participants also emphasized that the research community or individual reviewers were not necessarily bad or intentionally writing such reviews; instead, the Western-centric scholarly divide, tech-solutionist mindset, and reviewers' lack of lived experiences in the GS often led to unhelpful or dismissive reviews.} According to the participants we talked to, such reviews often came from researchers outside the GS, who often held rigid mental models of the region and tried to impose those on authors, dismissing alternative perspectives. For example, P16 raised concern about the prevalent power dynamics in knowledge evaluation, narrating:
\change{
\begin{quote}
\textit{``Sometimes I feel HCI conferences are wary of the words extraction, exploitation, etc, unless these issues are raised by the same community} [Global North scholars]''.
\end{quote}}

These insights suggest that structural inequities perpetuate WEIRD reviewing dynamics towards GS scholarship. 

\subsection{Experiences of Reviewing GS Scholarship}\label{sec:giving}
All our participants reported serving as reviewers \change{(refer to Table~\ref{tab:review_exp})}, while several served as area chairs, associate chairs, paper and program chairs for HCI venues. Their scholarly identity in relation to the GS frequently shaped what kinds of papers they were assigned to review, how they reviewed those work, and their interaction with fellow reviewers, \change{i.e., other reviewers assigned to the same paper in the double-anonymized review process.} 

\subsubsection{Who Gets to Review?}
Participants discussed various systemic issues in how reviewers are assigned in current peer review system to evaluate HCI research from the GS. 

\paragraph{Tokenization as A ``Cultural Expert''} Participants were frequently assigned to review papers that either focused on their home countries or GS in general, even when those papers fell outside their research area. P8 felt being \textit{``tokenized''}, i.e., valued more for their cultural identity rather than scholarly expertise. P4 described: 
\change{
\begin{quote}
\textit{``They} [AC] \textit{invited me to review a paper. That paper has nothing to do with my expertise in terms of the computing domain. The only thing that overlaps with me is that the paper is based in Bangladesh.''}
\end{quote}
}

Participants critically reflected that due to their privileged affiliations with Global North institutions, they were now treated as a \textit{``cultural expert''} for studies in the GS even when they did not share those cultural identities. For example, P4 often received requests to review papers on Muslim communities in Bangladesh, despite not belonging to that religious group and thus feeling unqualified to review them adequately. 

Additionally, participants reported facing ethical dilemmas when they were asked to review work about their home countries. Since HCI research communities in these regions were often small, sometimes participants could infer authors’ identities from the paper’s topic. When participants attempted to raise these potential conflicts of interest with associate chairs (ACs), their concerns were often dismissed. In some cases, they later discovered that the papers they had advocated for during the review process were, in fact, authored by friends or peers.

\paragraph{Assumptions About \change{Fellow} Reviewers' Identity} 
When the \change{fellow} reviewers overlooked or misunderstood the cultural context of a paper, participants often assumed they were from WEIRD regions. 
In such situations, participants mentioned taking on additional labor during the reviewer discussion period, 
having to explain and defend the cultural context of the GS research. 
\change{However, participants who were associate chairs and had greater visibility into reviewers' identities noted that WEIRD reviews could also come from GS reviewers.} P9 described:
\begin{quote}
    \textit{``I've actually had more arguments with other people who are also doing research for the Global Majority, or in the Global Majority$\dots$I don't think that it's necessarily always reviewers from the Global North who are stifling some of this research, because that's not the case.''}
\end{quote}

These findings \change{indicate that WEIRD reviewing dynamics are often systemic and cannot be reduced to North-South binaries or reviewer identity.} They highlight the need to improve the reviewer assignment process that moves beyond tokenizing cultural identity, ensuring that GS research is evaluated by reviewers with relevant expertise while reducing epistemic burden on fellow GS reviewers.

\subsubsection{How Is Reviewing Done?}
Participants highlighted several issues and asymmetries in current reviewing practices that disproportionately affect GS scholarship.

\paragraph{Hidden Curriculum in Writing} 
While serving as reviewers for HCI venues, participants noticed that critiques of GS research were not always about the content or rigor of the study. They observed that their \change{fellow} reviewers often tried to reject work due to writing format and structure. 
P13 stressed that there was a \textit{``hidden curriculum''} of writing for HCI venues, which normalized a Western-centric, objective, and scientific rationale form of writing. Very often, these norms were inaccessible to authors conducting research from the GS. 
Participants viewed this as a gate-keeping mechanism, excluding non-normative writing from the GS. 
\change{P7 noted that, as a scholar from the GS, they often faced the tension of reinforcing Western writing standards when they rejected work from the GS that deviated from these norms. 
P4 justified such practice saying that when papers did not adhere to established writing norms, Global North readers might view the publication as low-quality, 
perpetuating negative perceptions about research from the GS.} 

These dynamics suggest that the systemic prioritization of Western-centric writing norms in HCI, along with inequitable access to knowledge about these norms, reduces the diversity of voices and participation from GS researchers.

\paragraph{Resisting WEIRD Assumptions as Reviewers}
Participants reported several considerations they took into account when reviewing research from 
the GS. P1 talked about approaching papers from the GS with a mindset of \textit{``what makes the paper a get-in rather than a get-out of the conference?''}. Participants drew from their personal experiences of receiving WEIRD reviews and made a conscious effort to avoid writing such reviews. 
P10 emphasized:
\change{
\begin{quote}
``[B]\textit{ecause I'm already triggered by my own experience$\dots$as a very early career person receiving some of the feedback that wasn't really a feedback.''}
\end{quote}
}

Participants emphasized that, when writing reviews, they sought to trust the authors and acknowledge their agency in representing the situated knowledge and cultural contexts of  the study based in the GS. 
Further, considering the hidden curriculum in writing, P3 and P13 shared that they tried to provide detailed writing feedback to authors so that they could structure their papers effectively for the HCI audience. P7 noted that new researchers from the GS might be unaware of citation politics in HCI. 
To support these authors, participants provided review feedback on how to situate the paper within HCI literature. Participants also indicated utilizing rebuttal and reviewer discussion periods to challenge WEIRD assumptions made by \change{fellow} reviewers, 
with mixed outcomes---sometimes successful, other times not. 
They particularly reported calling out \change{fellow} reviewers, who made problematic remarks about the studied GS community or the author's identity instead of the merit of the work.
P10 shared:
\begin{quote}
    \textit{``I show up ready to fight during the discussion phase. I am like super engaged as a reviewer and as an AC because I do not want to see what happened to me being replicated. I check with reviewers to see if they meant XYZ because that's how the reviews may come across. But doing this takes time off my other responsibility.''}
\end{quote}

These findings highlight how GS researchers often bear the additional burden of serving both as \textit{cultural experts} to assess GS scholarship and resisting WEIRD assumptions against these work during the review process.

\subsection{Moving Beyond WEIRD Reviews}\label{sec:forward}
Our participants shared their strategies of how they tried to move past WEIRD reviews as authors and provided suggestions for systemic changes within the HCI community.

\subsubsection{How Do Authors Process WEIRD Reviews?} Participants developed several strategies over the years to cope with WEIRD reviews. 
As paper authors, they suggested processing biased reviews in phases, like taking a first pass to deal with the emotional toll, then taking subsequent passes to extract actionable steps. 
They emphasized the importance of engaging with reviewers during the rebuttal and author response periods. Overall, three strategies emerged from their suggestion: 1) ignoring the reviewer and biased review, 2) gently pushing the reviewer and explaining the research context explicitly, and 3) calling out problematic remarks by the reviewers. 

Senior participants shared that they received mixed outcomes when trying to assert their scholarly decisions during the rebuttal/ revision period. \change{Although senior participants 
could often afford resisting biased reviews, they recognized junior researchers might not have the same privilege. P9 stated:
\begin{quote}
\textit{``For me, it's how you move forward with any review but I'm white, privileged, and$\dots$senior. I have published so many papers that it doesn't matter much to me right now. But young researchers in my team are in need of publishing in top venues to make their presence known to the community. I do spend time with them talking through the reviews, suggesting what bits they can just ignore when it comes from reviewer's bias against the Global Majority. And then what they can fight about, what they can take} [from reviews] \textit{and change} [in their paper]. \textit{Because often reviews might not be fully biased but they misunderstand you.''} 
\end{quote}} 


Participants suggested relying on allies and collaborators, especially to get help with the writing, editing, and framing of the paper. 
For instance, P8 commented, ``\textit{You need an advisor who has published at CHI to publish at CHI}'' or P7 remarked, \textit{``you need a White ally to advocate for your paper''}---reflecting on the power dynamics and systemic hierarchy that GS researchers often have to navigate to get their work published. They urged allies with geographic, seniority, and social privileges to \textit{``speak up''} so that reviewers could be made aware when their approach was problematic. They argued this could create a learning opportunity for reviewers without putting less-privileged scholars at risk for challenging WEIRD reviews. 

\subsubsection{How to Constructively Review GS Research?}
Although participants acknowledged the voluntary labor of reviewers, they stressed that conscious efforts were necessary on the reviewer's end to improve the status quo. They recommended avoiding \textit{``high resource expectations''} (P4, P5) 
and recognizing the value of \textit{``situated knowledge''} (P5, P9, P14) 
when evaluating research from GS contexts. 
They suggested 
reviewers 
to be mindful of their own biases. They emphasized that any critique should be targeted at the rigor of the work rather than making presumptions about the authors or 
the studied community. 
Additionally, participants suggested that if the reviewer is rejecting work due to writing structure, they should provide 
detailed feedback for framing the work, since the hidden norms of writing an HCI paper might not be otherwise accessible to many GS authors. 
They also recommended that reviewers describe their expertise 
when submitting reviews, so that the area/ associate chair could make an informed decision. P8 explained: 

\begin{quote}
    \textit{``$\dots$when I'm possibly being culturally tokenized} [as a reviewer] \textit{for this kind of papers.$\dots$I'm coming off as someone who is familiar with the cultural context, social complexities, but I might not be familiar with the topic of the paper like, fintech in Bangladesh, particularly. so I'll be very transparent about that.''}
\end{quote}

Further, participants reflected on the role of area and associate chairs in addressing WEIRD reviews, resolving conflicts, and engaging co-reviewers in the discussion period, and going beyond simply summarizing reviews. P3 stated:

\begin{quote}
    \textit{``A meta reviewer should at least decide what feedback they think is justified, and what they think is not justified. That's a prime responsibility, I think, of a meta reviewer, is to evaluate all the reviews, and not just summarize the reviews provided by the reviewers and just give it back to the authors.''}
\end{quote}

\subsubsection{What Systemic Changes Are Needed?}
Participants pointed out that one-off efforts 
might not be effective until 
conferences deliberately make systemic changes to create truly inclusive spaces for GS scholarship. P2 remarked:

\begin{quote}
    \textit{``I think there is a much broader need of globalizing within HCI that isn't happening, right? Like, even the fact that CHI is happening in Barcelona, and next in Pittsburgh$\dots$So we're not going anywhere in the Global South just from the conference's standpoint, let alone supporting scholarship from the Global South.''}
\end{quote}

\paragraph{Formalizing Reviewer Selection}
Since reviews often pushed GS authors to adopt data-driven methods over community-centered ethnographic approaches, participants recommended appointing senior researchers with mixed-methods expertise as technical reviewers. 
For each paper, P3 suggested assigning one reviewer for domain expertise, one for cultural context, and one senior reviewer. They also advised formalizing the role of cultural expert, since these reviewers often undertake additional labor when \change{fellow} reviewers or domain experts misunderstand the cultural nuance. 

\paragraph{Improving Incentives and Learning Opportunity for Reviewers}
Participants noted that due to the shortage of qualified reviewers with the growth of conferences, reviewing workload disproportionately falls on junior researchers, who might not be familiar with the historical context of HCI research involving GS communities. They pointed out that the current system offers little incentive for senior academics to participate. Participants recommended reevaluating incentive structures to better engage both senior reviewers and train junior reviewers. 

Acknowledging that it might be difficult to hold in-person meetings of Program Committee (PC) with the current conference sizes, P5 and P9 proposed organizing virtual learning spaces or holding PC meetings during the conference so that senior researchers can share their reviewing experiences and expertise with junior researchers. Moreover, given the limited recognition of \textit{``invisible volunteer labor''} 
in reviewing, P5 and P9 proposed awards such as an \textit{``Inclusive Reviewing Award''}, 
particularly to encourage equitable and thoughtful review practices.

\paragraph{Codifying Reviewing Standards}
Participants emphasized the need to hold reviewers accountable through uniform reviewing policies, both for GN and GS research studies. For example, requiring all submissions to include positionality statements, generalize findings, or include the country name in the paper title. 
To prevent reviews that devalue GS scholarship based on implicit signals of authors' identity, such as grammar, P1 proposed that HCI conferences can learn from NLP venues 
like ACL or *CL, which prohibit commenting on the grammar of the paper. 
P2 also recommended adopting an open review policy followed by some machine learning conferences, arguing that 
making anonymous reviews publicly available would promote accountability and discourage problematic remarks.

Achieving equitable review practices in HCI is not the responsibility of individual authors or reviewers, but demands collective effort and systemic change within the community.

\section{Discussion}
Prior work has shown that GS scholarship is subjected to disproportionate scrutiny, such as the pressure to justify the novelty and generalizability of the work~\cite{Kumar-2021, soden2024evaluating, bender2019benderrule}, requests to specify locations in paper titles~\cite{Kou2018} and include positionality statements~\cite{Singh-2025}, paternalistic comments towards the studied GS communities~\cite{Kumar-2021}, and criticism of authors' English writing~\cite{Lepp-2025, Kumar-2021, herrera1999language, shchemeleva2021there}. Our study advances this conversation in three key ways.

First, we uncover diverse manifestations of WEIRD reviews that are yet to be systematically documented, including the dismissal of GS researchers’ theoretical contributions in HCI (restricting their expertise to ``development research''), \change{questioning researchers' credibility based on positionality statements}, and the assertion of monolithic assumptions about the GS to undermine researchers' expertise and situated findings. \change{While positivist critiques of qualitative and interpretive studies are well-documented in HCI peer review~\cite{soden2024evaluating}, our findings show that such reviews put additional burden on qualitative researchers studying GS contexts--forcing them to justify their epistemic sensibilities against prescriptive, data-driven methods that often misalign with local realities.}

Second, in contrast to prior work~\cite[e.g.,][]{De-2025, soden2024evaluating, bol2023address, amarante2022marginalization}, our study accounts for reviewers’ perspectives and reveals the tensions reviewers face when evaluating GS scholarship. We demonstrate \change{how GS HCI researchers are often tokenized as \textit{``cultural experts''} during peer review and how GS reviewers affiliated with Western institutions may unintentionally reinforce Western-centric writing norms that are often inaccessible to fellow researchers based in the GS.} We reveal the \textit{``epistemic burden''} experienced by our participants, both as authors and reviewers of GS scholarship, as they constantly have to explain why knowledge and cultural contexts of these communities matter.
Finally, our work moves beyond diagnosis by offering concrete recommendations from participants to improve peer review for GS scholarship in HCI. These include allyship from authors with geographic and seniority privileges, conscious efforts from reviewers and area/associate chairs to counter problematic reviews, and systematic reforms within HCI to codify protections against discriminatory review practices.

Drawing on our findings, we argue that the current peer review system perpetuates epistemic oppression towards GS scholarship, which is already marginalized within HCI~\cite{Linxen-2021, Kumar-2021}. While prior scholarship has applied epistemic oppression to critique institutional devaluation of knowledge produced by marginalized scholars~\cite{collins2017intersectionality, settles2021epistemic, catala2024academic, settles2024epistemic}, our work uniquely extends this framework to the peer review process itself, showing how it can operate as a gatekeeping mechanism that often undermines GS scholarship (Sec. \ref{sec:epistemic}). \change{As our participants emphasized, these reviews may not always be intentional, but the harms, labor, and burden they experience are nonetheless real, recurring, and consequential.} Importantly, these experiences reflect deeper systemic conditions that marginalize GS within academic knowledge production~\cite{moletsane2015whose, schopf2020conversations, collyer2018, sharma2025sustainability}. Building on this framing, we propose concrete steps toward \textit{unWEIRDing peer review in HCI}, grounding our recommendations in \citet{dotson2014conceptualizing}’s model of first-, second-, and third-order changes required to counteract epistemic oppression (Sec. \ref{sec:recs}).

\subsection{Epistemic Oppression in Peer review}\label{sec:epistemic}

\subsubsection{First-Order Epistemic Exclusion} 
As stated by \citet{dotson2014conceptualizing}, first-order epistemic exclusion is a persistent compromise of one’s epistemic agency due to their epistemically disadvantaged social identity. In the case of peer review, such exclusions manifest through (often unconscious) identity-prejudices that withhold epistemic credibility from marginalized scholars~\cite{causevic2020centering, Kumar-2021}. 

Our participants noted that reviews often delegitimized their intellectual contributions, confining them to ``development research'' based on their GS research. This dynamic underscores how entrenched North–South power hierarchies continue to shape knowledge evaluation, even amid growing academic mobility and claims of a transnational scholarly community~\cite{collyer2018}.
Additionally, our participants highlighted that their expertise, situated knowledge, and theoretical contributions in HCI were often questioned and devalued in reviews, primarily due to their English skills and engagement with GS communities. This dynamic reflects a broader and longstanding prejudice in academic publishing, where authors' proficiency in English is unfairly treated as a proxy for intellectual rigor~\cite{herrera1999language, egger1997language, lee2013bias, lopez2015publish,mckinley2018conceptualizations, grzybowski2019language, gonzalez2024science, Lepp-2025}. \change{To assert credibility, participants described bearing a disproportionate burden that required them to include positionality statements when studying GS contexts, which they felt was rarely applied to work focusing on Western contexts. Despite this additional labor, they still faced reviews that dismissed their expertise and situated knowledge based on monolithic and often problematic assumptions about GS communities.} 

Although 
these exclusionary dynamics broadly stem from geospatial and social stratification in academic knowledge production that have historically favored Western scholars and institutions, sustaining center-periphery structures across disciplines where the Global North dominates the academic capital~\cite{demeter2020academic, obeng2019intellectual,  collyer2018, marginson2023hegemony, Safir-2025, sharma2025sustainability}. Thus, the intellectual hegemony of the North perpetuates certain normative assumptions, in which {``other''} logics of knowledge are seen as {``exotic, objectionable or downright crazy''}~\cite[p.~218]{connell2014using}. As articulated by our participants in Sec. \ref{sec:receiveing}, in peer review in HCI, these dynamics manifest through disparate standards of novelty, generalizability, \change{and validity of design recommendations. While qualitative studies in HCI are frequently expected to generalize findings~\cite{soden2024evaluating}, our participants encountered reviews that presumed design contributions grounded in GS communities could not be generalized. Paradoxically, when the needs of these communities aligned with those observed in Western contexts, reviews frequently dismissed the work as lacking novelty. And findings based on WEIRD samples are considered universally applicable~\cite{baber2003provincial, cheon2020usa, sharma2025sustainability}.}

\change{This discrepancy reflects entrenched epistemic practices in which the knowledge derived from certain geographies is rendered unintelligible or positioned as inherently inferior to results derived from Western contexts~\cite{castro2022north, grech2011recolonising, connell2020southern}.} Although framed as ``neutral'' criticism, such accounts can disproportionately harm marginalized individuals in their capacity as knowers~\cite{dotson2012cautionary}. Early career researchers in our study recounted the negative impacts of such WEIRD reviews, leaving them with a sense of powerlessness and the {``invisible labor''} of constantly having to justify their scholarship compared to Western scholars. 

\subsubsection{Second-Order Epistemic Exclusion} \citet{dotson2014conceptualizing} explains that second-order exclusion occurs when shared epistemic resources reflect the language and social imaginaries of the dominant group, thereby remaining opaque to those who have historically been considered epistemically less credible. 
Our findings underscore how such exclusion affects GS scholarship via hidden norms and privileged mentorship, often inaccessible to scholars from the GS.

As reviewers, our participants observed that the hidden curriculum of writing HCI papers disadvantaged GS researchers, \change{pointing out that submissions were often rejected--both by themselves and fellow reviewers--}when the writing diverged from Western-centric conventions of objectivity and scientific rationale (Sec. \ref{sec:giving}). They emphasized that successful CHI publications typically required mentorship from advisors already embedded in those publishing networks, underscoring how these dynamics remain opaque to many GS researchers. These practices are deeply embedded in Western-centric publication regimes, where scholarship that diverges from dominant writing norms is either viewed as lacking sincerity or as failing to satisfy the epistemic tastes of the journal~\cite{collyer2018, getahun2021academic}. 

Our participants highlighted that the reputational capital of elite HCI venues compelled them to submit their work to these venues despite the differential treatment they face as GS scholars (Sec. \ref{sec:receiveing}). Our findings also reveal how researchers working on GS scholarship faced added scrutiny if their research did not draw from dominant Western frameworks. This dynamic frequently positions Western academics as a credible audience for the produced knowledge~\cite{getahun2021academic, bol2023address} and the choice of this audience influences which interpretive tools are used to evaluate and disseminate new knowledge~\cite{bhakuni2021epistemic}. Thus, reviews that insist on building on Western-centric frameworks disadvantage GS scholarship because those lenses to interpret data do not embody how GS communities make sense of their situated knowledge and experiences~\cite{Safir-2025}. 

\subsubsection{Third-Order Epistemic Exclusion} According to \citet{dotson2014conceptualizing}, third-order exclusion arises when the dominant epistemic agent willfully ignores and questions the credibility of marginally situated epistemic resources in knowledge production. This refusal undermines a knower’s ability to contribute to shared epistemic resources within a community, thereby compromising their epistemic agency. 

In our study, participants described encountering such injustice in the review process, particularly when their ethnographic and community-centered approaches in the GS conflicted with reviewers’ expectations for data-driven, generalizable results (Sec. \ref{sec:receiveing}). 
These experiences extend beyond the broader marginalization of qualitative research by quantitative experts~\cite{soden2024evaluating}. They reveal a deeper epistemic exclusion, in which dominantly situated knowers systematically devalue epistemic tools, including methods, theories, and frameworks developed in non-Western or peripheral contexts~\cite{pohlhaus2012relational}. Participants reported that reviews frequently ignored pluralist perspectives when their work drew on regional theories, enforcing restrictive citational norms that privileged Western scholarship. Even when participants attempted to assert their epistemic agency during rebuttals or revisions, their arguments were often dismissed or negatively received, illustrating how power differentials in the peer-review process operate to silence marginalized epistemologies within HCI communities.

Taken together, our findings demonstrate that the current peer review process reproduces Western hegemony within the contemporary knowledge economy. It does so by ingenuously perpetuating identity prejudices against GS scholars as neutral feedback, overlooking the disproportionate influence of privileged mentorship and tacit knowledge about the hidden norms of HCI publishing, and privileging Western epistemic frameworks. \change{We argue that these dynamics are not merely a matter of inconsistent reviewing, shortage of expert reviewers, or increased reviewing workload due to expanding conference sizes. While some participants speculated that Western scholars disproportionately produce such WEIRD reviews and others noted that GS scholars also enact these practices, our analysis indicates that these problems are systemic--sustained through dominant epistemic traditions that govern how reviewers assess scholarship, regardless of their identities.} 

\subsection{Towards Epistemic Inclusion in Peer Review}\label{sec:recs}
As our findings highlight, the WEIRD dynamics in HCI peer review result from systemic choices that exclude and, at times, marginalize GS scholarship. Our intention with this paper is not to dismiss the value of peer review, but rather to foreground the concerns of marginalized GS scholars and initiate a conversation among authors, reviewers, and the HCI community at large. In this section, we build on our participants’ recommendations (Sec. \ref{sec:forward}) and apply the concept of epistemic oppression~\cite{dotson2014conceptualizing} to ground actionable steps toward cultivating a more inclusive and just scholarship, and, by extension, a community. \change{We first summarize our recommendations in Table~\ref{tab:recs} and provide additional details below.}

\begin{table*}[!ht]
\caption{Recommendations for improving reviewing experiences for Global South (GS) scholarship based on \citet{dotson2014conceptualizing}'s framework of epistemic oppression.}
\label{tab:recs}
\centering
\renewcommand{\arraystretch}{1.35}
\setlength{\dashlinedash}{1.5pt}
\setlength{\dashlinegap}{2pt}
\begin{tabular}{@{}c >{\centering\arraybackslash}m{2cm} p{12cm}@{}}
\toprule
\textbf{Framework} &
  \textbf{Stakeholders} &
  \textbf{Recommendations} \\ \midrule
\multirow{3}{*}{\begin{tabular}[c]{@{}c@{}}First-order\\ changes\end{tabular}} &
  Authors &
  \begin{tabular}[t]{@{}l@{}}$\bullet$ Gradually processing WEIRD reviews and the emotional toll\\ 
  $\bullet$ Utilizing rebuttal period to address WEIRD reviews\end{tabular} \\ \cdashline{2-3}
 &
  Allies &
  $\bullet$ Provide writing feedback to GS authors \\ \cdashline{2-3}
 &
  Reviewers &
  \begin{tabular}[t]{@{}l@{}}
  $\bullet$ Including reviewer positionality statements to report cultural expertise \\ 
  $\bullet$ Providing concrete writing and framing feedback to authors
  \end{tabular} \\ \midrule
\multirow{3}{*}{\begin{tabular}[c]{@{}c@{}}Second-order\\ changes\end{tabular}} &
  Reviewers &
  \begin{tabular}[t]{@{}l@{}}
  $\bullet$ Actively checking one's biases during review \\ 
  $\bullet$ Countering WEIRD feedback from fellow reviewers\end{tabular} \\ \cdashline{2-3}
 &
  Program committee &
  \begin{tabular}[t]{@{}l@{}}
  $\bullet$ Reporting the quality of reviews based on meta-reviewers' feedback \\ 
  $\bullet$ Reporting geographical breakdown of paper acceptance rates across subcommittees\end{tabular} \\ \cdashline{2-3}
 &
  Community &
  \begin{tabular}[t]{@{}l@{}}
  $\bullet$ Holding co-learning sessions for early-career researchers and first-time reviewers \\ 
  $\bullet$ Having open discussions about the use of LLMs in reviewing \\ 
  $\bullet$ Allowing junior scholars to shadow senior scholars' reviewing strategies \\ 
  $\bullet$ Holding workshops for GS authors to demystify how to write HCI papers \\
  $\bullet$ Documenting best practices and available resources to address WEIRD reviews
  \end{tabular} \\ \midrule
\multirow{2}{*}{\begin{tabular}[c]{@{}c@{}}Third-order\\ changes\end{tabular}} &
  Program committee &
  \begin{tabular}[t]{@{}l@{}}
  $\bullet$ Standardizing requirements of positionality statements for all\\ 
  $\bullet$ Prohibiting reviews requiring GS scholarship to add location name in paper titles \\ 
  $\bullet$ Adopting formal grievance and reporting mechanisms to address biased reviews \\ 
  $\bullet$ Requiring meta-reviewers to assess review quality\end{tabular} \\ \cdashline{2-3}
 &
  SIGCHI &
  \begin{tabular}[t]{@{}l@{}}
  $\bullet$ Ensuring conference registration discounts are reflective of the realities of GS\\ 
  $\bullet$ Diversifying conference locations \\ 
  $\bullet$ Partnering with regional chapters to facilitate engagement of GS scholars \end{tabular} \\ \bottomrule
\end{tabular}
\end{table*}


\subsubsection{First-Order Changes} 
\citet{dotson2014conceptualizing} describes first-order change as resolving problems as they arise while keeping the established norms in place. 
In dealing with WEIRD Reviews, first-order changes refer to problem-solving strategies that authors, reviewers, and allies can adopt within their own capacities. 

\paragraph{Authors and The Allies of GS Scholarship:} 
For scholars who are from the GS, it might often be difficult to extract valuable feedback when reviews target their identity. As our participants suggested, in such cases, authors might try processing the reviews in phases, taking time to recuperate from the emotional toll of such reviews (Sec. \ref{sec:forward}). 
WEIRD reviews could also be targeted towards the communities studied within GS scholarship. 
In these cases, scholars have a responsibility to stay true to the needs of the communities, while also critically reflecting on the privileges they hold in conducting such research~\cite{le2022confronting}.
Additionally, authors may utilize existing rebuttal and author response periods 
to counter the WEIRD reviews. However, the ability to use these mechanisms effectively often depends on the author's seniority, institutional support, and broader privileges, leaving some scholars less comfortable and less protected in doing so. Here, allies, such as mentors or peers with experiences of publishing in HCI, could help GS scholars navigate WEIRD reviews by providing writing feedback~\cite[e.g.,][]{Ming-2025} and sharing strategies for rebuttal~\cite[e.g.,][]{soden2024evaluating}. 
\change{However, these efforts at the individual level may not be feasible within limited peer review cycles. Hence, it is imperative to consider systemic changes to relieve the burden from individuals, which we discuss later in this section.}
\paragraph{Reviewers:} 
In the current peer review system, GS scholars are often assigned as ``cultural experts,'' resulting in a mismatch between their expertise and the paper's technical contributions (Sec. \ref{sec:giving}). Similarly, GN scholars may be assigned to review papers, where they may lack the cultural expertise to provide effective feedback. 
 While most reviewing platforms allow for self-reported expertise and confidence scores, these metrics might not adequately capture these dynamics. 
HCI reviewers could consider including ``reviewer positionality statements'' when evaluating papers to explicitly report their cultural knowledge. 
To alleviate the impact of ``hidden curriculum'' in writing HCI papers, reviewers could provide concrete feedback on how to structure the paper, pointing to citations that are foundational to the field but remain opaque to GS scholars. 
Additionally, reviewers with expertise in GS scholarship can utilize the reviewer discussion period to engage with \change{fellow} reviewers, who may misunderstand the cultural context of a study. 

There have already been promising efforts to make the review process less opaque and more inclusive. For example, some HCI scholars have co-authored a living document that serves as an ``unofficial'' guide to peer review~\cite{kumar2020_unofficial_seven_stages_review_chi}. These initiatives are commendable, particularly as they are grounded in care and solidarity, given that reviewing itself is an unpaid labor~\cite{El-Guebaly2022Sep, Riley2016Dec}. Building on these efforts, as HCI scholars, we could develop an official and periodically updated living document tailored to GS scholarship. 

\subsubsection{Second-Order Changes} 
This change occurs when ``sets of individuals or groups are willing to alter their values and thereby create new strategies or ways of thinking and acting that actually improve the effectiveness of the system''~\cite{walsh2004interpreting, dotson2014conceptualizing}. We describe how we can introduce such changes within the peer review system.

\paragraph{Altering the Status Quo:}
Our findings highlight how reviews often dismiss situated knowledge from the GS (Sec. \ref{sec:receiveing}).
One way to address this is to approach reviewing with curiosity, especially in contexts that require subjective interpretations.
When reviewing GS scholarship, reviewers need to play the role of an {``inquirer''}~\cite{Ritunnano2022Jul}, recognizing their contextual gap and actively asking questions that reflect the intent to reach a common ground. 
Reviewers could embrace the plurality of knowledge production and reconsider whether feedback to apply Western frameworks in GS contexts would be appropriate~\cite {Kambunga2023May, Mignolo2007Mar, Safir-2025}. Additionally, prior work in HCI has called for more explicit involvement from allies to dismantle racist and oppressive systems~\cite{Ogbonnaya-Ogburu2020Apr}. As a step in that direction, scholars reviewing GS scholarship should nurture {``effective allyship''}~\cite{Bhattacharyya2024May} that resists WEIRD reviews from \change{fellow} reviewers whenever possible. \change{However, as participants pointed out, the labor behind such resistance, future research can examine ways to develop sustainable approaches to better support allies and reviewers in addressing such biases.}

\paragraph{Enhancing Transparency around Publication Process:}
Most HCI venues have established mechanisms for transparency, routinely disclosing the status of submissions and publications. 
For example, CHI publicly releases a program committee report\footnote{https://chi2025.acm.org/chi-2025-papers-track-post-pc-outcomes-report/}, which provides a summary of the submitted and accepted papers, including a geographic breakdown of these submissions. These reports could go a step further to include a primary analysis of the quality of reviews that meta-reviewers often assign (e.g., highly useful, not useful), particularly for papers from different countries. However, not all meta-reviewers specify the quality of reviews, and formalizing this process would require conscious efforts from them. 
Additionally, these reports could give a breakdown of accepted papers--either from or about the GS--across different subcommittees or tracks, which could reveal if certain tracks are more receptive to diverse perspectives from the GS. These details could help identify what is working and what needs to change in current review practices.

\paragraph{Fostering Reviewer Co-Learning Spaces:}
Our participants iterated the need for creating formal learning spaces, such as \change{Program Committee (PC)} meetings, which could be beneficial for early-career researchers and first-time reviewers (Sec. \ref{sec:forward}). 
While written guidelines can serve as valuable resources, reviewing workload and commitments across different conferences often makes it difficult for reviewers to engage with such content without any particular incentive~\cite{Rogers2023Jul}.
Instead of passive, undirected resources, co-learning spaces that prioritize active engagement among reviewers may enable more effective knowledge sharing. 
Co-learning spaces would be especially beneficial in current times, when reviewers who feel uncertain about their expertise might otherwise turn to Large Language Models (LLMs) to evaluate a paper~\cite{Liang2024Jul}. While AI-generated reviews affect the entire scientific community~\cite{Kocak2025Jun}, these may be especially detrimental to GS scholarship as many LLMs produce stereotypical and misaligned interpretations of non-Western communities~\cite{Naous2024Aug}.
Senior scholars, who may be less likely to accept review requests~\cite{El-Guebaly2022Sep, CANDALPEDREIRA202354} could also partake in co-learning spaces to share their expertise with junior scholars. 
Senior scholars who delegate reviews to junior researchers or their advisees could initiate co-learning opportunities by allowing the junior scholar to shadow the reviewing strategies for a few papers before going on to review on their own~\cite{El-Guebaly2022Sep}. 
It is important to note that senior scholars are not exempt from writing WEIRD reviews. 
Hence, for all reviewers, shared co-learning spaces can foreground critical reflections on addressing implicit biases and active participation discussed above. 

\paragraph{Centering GS Experiences and Needs:}
Our findings suggest that GS scholars face disadvantage in peer review due to their limited access to epistemic resources (Sec. \ref{sec:epistemic}). As a scholarly community, we need to prioritize hosting learning sessions and workshops for GS scholars to help them with paper writing norms for HCI venues (\cite[e.g.][]{Ming-2025}) since writing a good CHI paper depends on ``knowing the quirks of CHI''~\cite{Deterding-2023}. Such sessions could be held at conferences themselves and accommodate virtual attendance to make these spaces accessible to GS scholars who are often unable to attend conferences, which are predominantly held in the West~\cite{Linxen-2021}. 
Additionally, some Western institutions have writing resources~\cite[]{CHIbootcamp_2025} which can be extended and tailored to GS communities. This is not to dismiss that GS scholars do not have their own epistemic resources~\cite{Lobo2022Dec}. Rather, it is a call to bring forth disjoint and isolated resources to the mainstream. Consequently, we need to foster safe knowledge sharing spaces (e.g., Global South Special Interest Group~\cite{Upadhyay-2024}, HCI across borders~\cite{Kumar-2017}), where GS scholars can share their grievances and experiences. Such sessions could help document best practices, take stock of available resources, and remove a sense of isolation for GS scholars. 

\subsubsection{Third-Order Changes} 
A third-order change involves recognizing and potentially enabling the transformation of established social imaginaries that shape and sustain the prejudiced status quo~\cite{dotson2014conceptualizing}. In our context, this would involve recognizing the systemic values and behavior that contribute to WEIRD reviews and possibly trying to address them.

\paragraph{Codifying Uniform Review Standards:} 
As described in Sec. \ref{sec:receiveing}, GS scholars often face requests to include positionality statements, specify locations in paper titles~\cite{Kou2018}, or ``over-explain'' cultural contexts compared to studies in Western contexts. While reactions to positionality statements vary~\cite[e.g.,][]{Davis2020May, King2024Apr, Ganga2006May}, such statements may reinforce gatekeeping against marginalized scholars, as reviewers may use them to validate preconceived notions about who is ``qualified'' to do certain types of research~\cite{liang2021embracing}. To address this inequity, conferences could standardize the requirement for all authors or leave the decision entirely to author discretion.

On the other hand, including the country name in the paper title or abstract may negatively impact GS scholarship, as such geographic references consistently lead to lower visibility and fewer citations~\cite{abramo2016effect}. In contrast, studies in Western contexts are rarely expected to mention geography, often presenting their findings as universal~\cite{baber2003provincial, cheon2020usa, castro2022north}. To minimize such epistemic harm, conferences need to strive to prohibit reviews that single out GS scholarship by requiring location names in titles. Additionally, our participants reflected that the requirement to describe cultural context often puts a burden on them when writing for venues with strict page (e.g., FAccT, AIES) or word limits (e.g., CHI). Conferences could standardize how much extra space authors would receive when required to explain cultural or geographic contexts. 



\paragraph{Formalizing Accountability and Reporting Procedure:}
The lack of accountability and formal mechanisms to address WEIRD reviews in HCI creates a self-perpetuating cycle. To alleviate this issue, our participants suggested adopting an open review policy, i.e., making reviews publicly available, with reviewers free to disclose their identities. While there are limitations to open review~\cite{nature-1999}, longitudinal studies in machine learning communities show that it can foster transparency and accountability in knowledge evaluation~\cite{yangposition}.

Moreover, HCI venues could establish formal grievance and reporting mechanisms, modeled after systems implemented in NLP communities such as ACL~\cite{rogers2020can}. These mechanisms would allow authors to anonymously report predefined forms of misconduct, including unprofessional or rude reviews, inappropriate remarks targeting authors’ identities or the studied populations, and unjustified demands for novelty, among others. Such systems would provide marginalized authors with a secure avenue to report problematic reviews without fear of retaliation. In addition, consistent with practices adopted in other conferences, HCI venues could implement measures to hold reviewers accountable when they are found to repeatedly produce problematic reviews~\cite{arr}. Although currently it is optional for meta-reviewers to rate review quality, conferences could make it mandatory to help ensure higher-quality reviews. If such a system is in place, disclosing review quality across papers from different geographies in the existing program committee reports could increase transparency. These policies centered around the review process could complement existing initiatives like SIGCHI CARES\footnote{https://sigchi.org/about/sigchi-cares/} that focus on supporting individuals in the aftermath of discriminatory incidents.


\paragraph{Beyond Peer Review}
Prior research has highlighted the underrepresentation of GS scholarship in HCI publications~\cite{Sturm-2015, Linxen-2021, septiandri2024}. Whereas these studies primarily ask, ``How WEIRD are HCI venues?'', our study shifts the focus to ``Why is HCI WEIRD?''---critically examining the peer review process as one of the several structural factors that disadvantage GS scholarship in HCI. Participants in our study stressed that the organization of major HCI conferences in Western countries reinforces these inequities (Sec. \ref{sec:forward}), as GS scholars often face additional barriers, including financial constraints and visa restrictions~\cite{lewis2024conferencing, Dogan2023}. While there are efforts to increase the participation of GS scholars, for instance, through discounted registration rates, they are not reflective of the realities of the GS scholars and account for only a fraction of the overall cost to attend an international conference~\cite{Affairs2022Aug}. 

While regional initiatives such as AfriCHI~\cite{africhi2016Nov}, ArabHCI~\cite{Shaimaa2023Sep}, HCI4SouthAsia~\cite{hci4SA}, and Asian CHI~\cite{asia-chi} provide important avenues for knowledge exchange, participants noted that these venues often carry lower prestige relative to elite HCI conferences and offer limited career payoffs, which discouraged them from publishing in local chapters. This pattern reflects broader disparities in power and prestige between GN and GS publication venues~\cite{collyer2018}. Beyond diversifying conference locations, HCI venues might consider partnering with local chapters to facilitate scholarly engagement opportunities for GS researchers, who face financial or logistical constraints that limit their ability to attend conferences in Western countries~\cite{sharma2025sustainability}.

Thus, ensuring a fair peer-review process for the GS scholarship requires changes at multiple levels, from the roles and actions of individual actors to systemic changes that entail significant reconsideration of existing practices. While we have put forth recommendations based on our focus group discussions, we recognize these might not be exhaustive in addressing the epistemic harms faced by diverse GS scholars. Rather, we offer them as starting points to spark dialogue and explore potential pathways that the community can collectively debate, refine, and expand. 

\section{Conclusion: An Invitation}
Although peer review legitimizes scholarship, it often perpetuates biased reviews, particularly against underrepresented GS scholars and their work. In HCI, researchers have begun to interrogate epistemological biases, but review biases have not yet received direct attention. To address this gap, we conducted focus group discussions with HCI researchers studying GS contexts and analyzed their accounts through the lens of epistemic oppression. Participants described challenges such as the epistemic burden of repeatedly justifying why knowledge from GS communities matters, among others. Building on these insights, we propose first-, second-, and third-order changes to mitigate the harms of well-intentioned but structurally biased reviews and cultivate more equitable practices of scholarly evaluation.

The work to address review biases does not end with this paper; it is an ongoing effort that requires collective action, love, and care. We invite the SIGCHI community to join us in this effort, working independently and together to make HCI a more globally inclusive field. This requires recognizing the systemic inequities embedded in our review and publishing practices as well as reimagining them. By cultivating critical reflexivity about our individual actions and the scholarly norms, building accountability into peer review, and amplifying the voices of GS scholars and diverse underrepresented contexts, we can work toward creating an inclusive research community where plural epistemologies are valued and sustained. 
As Statistician C. Genest~\cite{genest1993comment} reminds us: 

\begin{quote}
    ``There is a long tradition attached to the peer review system. As users of science, we all depend on it: our professional realizations are based upon the work of others, and we count on journal (and book) editors to separate the wheat from the tares$\dots$As producers of science, it is also in our interest that the system be fair: favoritism, discrimination, and condescension bring discredit on the entire operation and ultimately work against the discipline, even if individual benefits occasionally may accrue in the short term.''
\end{quote}

To conclude, inclusivity and justice in peer review should not be treated as an afterthought; rather they need to be considered and acted upon from the very beginning. 


\bibliographystyle{ACM-Reference-Format}
\bibliography{8_references}


\end{document}